# Stochastic Optimization of Coupled Power Distribution-Urban Transportation Network Operations with Autonomous Mobility on Demand Systems

Han Wang, *Member, IEEE*, Xiaoyuan Xu, *Member, IEEE*, Yue Chen, *Member, IEEE*, Zheng Yan, Mohammad Shahidehpour, *Life Fellow, IEEE*, Jiaqi Li, Shaolun Xu

*Abstract*—Autonomous mobility on demand systems (AMoDS) will significantly affect the operation of coupled power distribution-urban transportation networks (PTNs) by the optimal dispatch of electric vehicles (EVs). This paper proposes an uncertainty method to analyze the operational states of PTNs with AMoDS. First, a PTN operation framework is designed considering the controllable EVs dispatched by AMoDS as well as the uncontrollable driving behaviors of other vehicle users. Then, a bi-level power-traffic flow (PTF) model is proposed to characterize the interaction of power distribution networks (PDNs) and urban transportation networks (UTNs). In the upper level, a social optimum model is established to minimize the operating cost of PDNs and UTNs embedded with controllable EVs. In the lower level, a stochastic user equilibrium (SUE) model is established to minimize the operating cost of uncontrollable EVs and gasoline vehicles (GVs) in UTNs. Finally, a probabilistic PTF analysis method is developed to evaluate PTN operations under environmental and human uncertainties. A regional sensitivity analysis method is proposed to identify the critical uncertainties and quantify the impacts of their distribution ranges on PTN operations. The effectiveness of the proposed method is verified by the PTN consisting of a 21-bus PDN and a 20-node UTN.

## NOMENCLATURE

### A. Acronyms:

| | |
|---|---|
| EV | Electric vehicle |
| PDN | Power distribution network |
| UTN | Urban transportation network |
| CS | Charging station |
| PTN | Power distribution-transportation network |
| UE | User equilibrium |
| SO | Social optimum |
| GV | Gasoline vehicle |
| SOC | Second-order cone |
| AMoDS | Autonomous mobility on demand systems |
| PTF | Power-traffic flow |
| SUE | Stochastic user equilibrium |
| RDG | Renewable distributed generator |
| RSA | Regional sensitivity analysis |
| LRA | Low-rank approximation |
| LMP | Locational marginal price |
| PDF | Probability density function |
| CDF | Cumulative distribution function |
| MCS | Monte Carlo simulation |

This work was supported by the National Natural Science Foundation of China under Grants (No. 52107116, U2166201). H. Wang, X. Xu, Z. Yan, J. Li, and S. Xu are with the School of Electronic Information and Electrical Engineering, Shanghai Jiao Tong University in Shanghai, 200240, China (e-mail: wanghan9894@sjtu.edu.cn; xuxiaoyuan@sjtu.edu.cn; yanz@sjtu.edu.cn; ljq0324@sjtu.edu.cn; slxu@sjtu.edu.cn). Y. Chen is with the Department of Mechanical and Automation Engineering, the Chinese University of Hong Kong, HKSAR, China. (email: yuechen@mae.cuhk.edu.hk). M. Shahidehpour is with the Electrical and Computer Engineering Department, Illinois Institute of Technology in Chicago (E-mail: ms@iit.edu).

### B. Parameters:

| | |
|---|---|
| $\lambda$ | Electricity price at slack bus |
| $\gamma$ | Penalty cost coefficient |
| $w$ | Cost coefficient for travel time |
| $E_C$ | Charging demand of each EV |
| $P_j^G$ | Active power output of traditional DG at bus $j$ |
| $Q_j^G$ | Reactive power of traditional DG at bus $j$ |
| $P_j^R$ | Active power of RDG at bus $j$ |
| $Q_j^R$ | Reactive power of RDG at bus $j$ |
| $r_{ij}^l$ | Resistance of line $l$ connecting buses $i$ and $j$ |
| $x_{ij}^l$ | Reactance of line $l$ connecting buses $i$ and $j$ |
| $P_j^D$ | Active power demand at bus $j$ |
| $Q_j^D$ | Reactive power demand at bus $j$ |
| $z_{ij}^l$ | Impedance of line $l$ |
| $i_l^{\max}$ | Limit for $i_{ij}^l$ |
| $U_i^{\max}$ | Upper bound of $U_i$ |
| $U_i^{\min}$ | Lower bound of $U_i$ |
| $P_i^{G,\max}$ | Maximum active power generation at bus $i$ |
| $P_i^{G,\min}$ | Minimum active power generation at bus $i$ |
| $Q_i^{G,\max}$ | Maximum reactive power generation at bus $i$ |
| $Q_i^{G,\min}$ | Minimum reactive power generation at bus $i$ |
| $P_j^T$ | Base load demand at bus $j$ |
| $P_j^{R,\max}$ | Maximum active power of RDG at bus $j$ |
| $q_{rs}^A$ | Actual travel demand for controllable EVs |
| $q_s^P$ | Prospective travel demand for controllable EVs at destination $s$ |
| $q_r^P$ | Idle EVs for rebalance travel at origin $r$ |
| $X_a$ | Limit of EV flow on charging link $a$ |
| $t_a^0$ | Free travel time of vehicles on link $a$ |
| $C_a$ | Capacity of link $a$ |
| $q_{rs}^G$ | Travel demand of GVs |
| $q_{rs}^E$ | Travel demand for uncontrollable EVs |
| $N_{in}$ | Sample size |

### C. Variables:

| | |
|---|---|
| $P_{0k}^l$ | Active power flow through line $l$ from bus 0 to bus $k$ |
| $P_{ij}^l$ | Active power flow through line $l$ from bus $i$ to bus $j$ |
| $Q_{ij}^l$ | Reactive power flow through line $l$ from bus $i$ to bus $j$ |
| $U_i$ | Squared voltage magnitude of bus $i$ |
| $i_{ij}^l$ | Squared current magnitude through line $l$ from bus $i$ to bus $j$ |
| $P_j^E$ | EV charging demand on charging link $a$ connecting to bus $j$ of PDN |
| $x_a^U$ | Controllable EV flow passing link $a$ |
| $x_a^L$ | Uncontrollable EV flow passing link $a$ |
| $\Delta P_j^R$ | Active power curtailment of RDG at bus $j$ |
| $f_k^{rs,A}$ | Controllable EV flow on path $k$ from $r$ to $s$ for satisfying actual travel demands |
| $f_k^{rs,P}$ | Controllable EV flow on path $k$ from $r$ to $s$ for satisfying prospective travel demands |
| $x_a^{U,A}$ | Controllable EV flow on link $a$ for satisfying actual travel demands |
| $x_a^{U,P}$ | Controllable EV flow on link $a$ for satisfying prospective travel demands |
| $x_a^U$ | Controllable EV flow on link $a$ |
| $x_a^L$ | Traffic flow of GVs and uncontrollable EVs on link $a$ |
| $t_a$ | Travel time on link $a$ |
| $f_k^{rs,G}$ | Traffic flow on path $k$ from $r$ to $s$ for satisfying travel demand of GVs |
| $f_k^{rs,E}$ | Traffic flow on path $k$ from $r$ to $s$ for satisfying travel demand of EVs |
| $x_a^{L,G}$ | GV flow on link $a$ |
| $x_a^{L,E}$ | Uncontrollable EV flow on link $a$ |
| $n$ | Dimension of input variables |
| D. **Sets:** | |
| $\pi(j)$ | Set of children buses of bus $j$ in PDN |
| $S_r$ | Set of origins |
| $S_s$ | Set of destinations |
| $K_{rs}$ | Set of paths connecting O-D pair $r$-$s$ |
| $S_C$ | Set of charging links |

I. INTRODUCTION

ELECTRIC vehicles (EVs) are becoming a promising alternative to reduce greenhouse gas emission and embrace green electricity [1]-[3]. With the proliferation of EVs, the interaction between power distribution networks (PDNs) and urban transportation networks (UTNs) is greatly strengthened [4]. On the one hand, traffic congestion could influence the EV users' decision for selecting charging stations (CSs), thus affecting the PDN electricity demands. On the other hand, the corresponding PDN operation could determine the price of electricity supplied to CSs. The EV users will take charging costs into account when making travel decisions to specific CSs, which in turn affects UTN states. The interdependence of PDNs and UTNs makes it challenging to depict and perceive the operational state of coupled power distribution-urban transportation networks (PTNs) [5]-[7].

The user equilibrium (UE) and social optimum (SO) [8] principles are usually used to model PTN operations. Under the UE principle, EV users determine travel routes to CSs by minimizing their travel costs. The UE principle characterizes EV users' rationality and has been widely applied in existing studies [9]-[13]. In [9], a PTN equilibrium model is established based on the UE principle. In [10], various UE models are revisited, and the interdependence between PDNs and UTNs is highlighted. In [11] and [12], the UE principle is embedded into a power-traffic flow model, and the interactive impacts of PDNs and UTNs are analyzed. In [13], a variational inequality approach is proposed to study the PTN pricing problem using a mixed UE model. In [14], a generalized UE model is proposed to investigate the coordinated PTN operation. The SO principle aims to improve the total social welfare by controlling EV driving behaviors and coordinating all PTN resources. In [15], an optimal pricing model is established to minimize PDN power losses and UTN travel costs. In [16], a collaborative scheme for PDNs and UTNs is proposed to make the system approach an SO status. In [17], a second-order cone (SOC) programming model is established to minimize the total PTN operating cost.

Compared with UE, SO is an ideal paradigm for controlling the EV user's driving behavior [10]. With the development of artificial intelligence, the EV self-driving techniques will become more readily available [18], which makes it possible to achieve SO by controlling EVs in an autonomous mobility on demand systems (AMoDS) [19]. Ref. [20] reviews the shared mobility-on-demand systems, where EV fleets are operated by a centralized platform. In [21] and [22], the interaction between AMoDS and PDNs is investigated. In [23], a strategic charging pricing scheme for AMoDS is designed based on reinforcement learning.

These studies investigate the PTN operation considering the impact of AMoDS, while neglecting other types of vehicles, such as private EVs and gasoline vehicles (GVs). In PDNs, EV charging demands are more variable than those of self-driving EV fleets controlled by AMoDS. In UTNs, driving behaviors of EVs and GVs are selfish and to a great extent uncontrollable, which could dramatically affect the traffic flow assignments. Hence, it is still an open issue to establish a PTN model considering complex vehicle flows and different driving behaviors. In addition, existing studies usually ignore the uncertainties associated with PTN operations. The variable power outputs of renewable distributed generators (RDGs) could affect the PDN supply and electricity prices. In UTNs, random traffic congestion could increase the traveling, causing EV users to change their routes and CS choices. Therefore, it is necessary to quantify the impacts of uncertainties on PTN operations. To address these issues, this paper proposes a PTN uncertainty analysis method with the integration of AMoDS. The main contributions of the paper are summarized as follows:

1. A PTN operation framework with AMoDS is proposed to characterize the interactions of PDNs and UTNs by considering controllable and uncontrollable driving behaviors. A bi-level power-traffic flow (PTF) model is established to depict PTN operations. In the upper level, an SO model is built to minimize the operating costs of PDNs and UTNs, which are embedded in controllable EV operations. In the lower level, the stochastic path choice behaviors of vehicle users are depicted using a logit model, and a stochastic UE (SUE) model is utilized to describe the UTN traffic equilibrium with uncontrollable EVs and GVs. In contrast to the existing studies [9]-[17], the proposed bi-level model incorporates different driving behaviors in vehicle flows.

2. A probabilistic PTF analysis method is developed to consider PTN operational states under uncertainties. In addition to variable power outputs of RDGs, which are commonly seen in the existing works, the variations of UTN link capacities, travel demands, and path selections of uncontrollable vehicle users are investigated in the probabilistic PTF analysis model.
3. A regional sensitivity analysis (RSA) method is proposed to quantify the impacts of uncertainties on PTN operations. A Copula-based $\delta$ index is derived to identify critical uncertainties, and a regional sensitivity index is introduced to quantify the impacts of variation ranges of uncertain inputs. Also, a low-rank approximation (LRA)-based surrogate model is built to improve the computational efficiency of sensitivity calculations.

The rest of the paper is organized as follows: The operation framework and the bi-level PTF model for PTNs with AMoDS are given in Section II. The probabilistic PTF analysis method and the RSA method are developed in Sections III and IV, respectively. Simulation results are given in Section V, followed by the conclusion in Section VI.

## II. Proposed Problem Formulation

In this section, a PTN operation framework with AMoDS is proposed to analyze the interaction of PDNs and UTNs. Then, a bi-level PTF model is established to describe the PTN operations with AMoDS.

### A. Operation Framework for PTN with AMoDS

Large-scale EV integration increases the interaction of PDNs and UTNs. To study the impacts of EVs, we divide the EV driving behaviors into the following two types:
1. *Controllable driving behaviors:* EVs obtain orders from the system with AMoDS and adjust their behaviors including order-serving, vehicle rebalancing, and charging [20]. In this regard, EV driving behaviors are considered controllable, especially when considering aided-driving and self-driving [23]. In this paper, we assume that EVs scheduled by AMoDS are controllable.
2. *Uncontrollable driving behaviors:* In general, EV driving behaviors are determined by individual rationality. Various factors can affect EV driving behaviors, such as personal preferences, and the cognitive bias regarding the travel time and cost. Hence, EV driving behaviors are considered uncontrollable.

Considering the two types of EV driving behaviors, we propose in Fig. 1 a PTN operation framework with AMoDS. For the controllable EV driving behaviors, AMoDS obtain the information on PTN operations (e.g., PDN electricity prices and UTN traffic flows), and release the EV control strategies. Hence, AMoDS can optimize EV travel routes and assist PTN operators to realize SO. For the uncontrollable EV driving behaviors, drivers estimate their charging and travel costs, and determine travel routes to designated CSs under the UE principle. Both controllable and uncontrollable EV driving behaviors would affect the PTN operations and cause complex interactions of PDNs and UTNs. We propose a bi-level PTF model in the following sections to depict the PTN operational state with AMoDS.

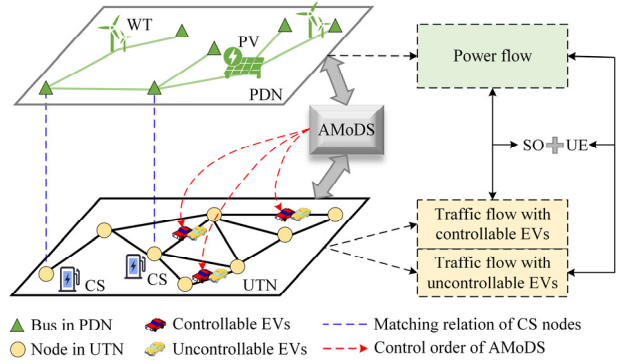

Fig. 1. Proposed operation framework for PTN with AMoDS.

### B. Proposed Model Assumptions

The corresponding assumptions are given as follows:
1. We assume that the EV driving behaviors are uncontrollable and their behaviors obey the UE principle.
2. We assume AMoDS are operated by independent operators to control shared EVs. The AMoDS signs contracts with PDNs to lower EV electricity prices.
3. We adopt a constant coefficient to represent the relationship between travel cost and time. In practice, the coefficient could depend on EV users.
4. We assume all EVs would need to be charged once during the trip. Otherwise, EVs would be treated as GVs. Also, GV driving behaviors are considered uncontrollable in this paper.

### C. Proposed Power-Traffic Flow Model

We establish a bi-level PTF model to describe the PTN operation with AMoDS. The bi-level model includes an SO model in the upper level and an SUE model in the lower level. The upper-level model minimizes the total PTN operating cost with controllable EVs. The lower-level model aims to achieve a network equilibrium considering uncontrollable EVs and GVs.

*1) Upper-level SO Model:* A PDN is depicted by a graph $G_P = (S_N, S_L)$, where $S_N$ denotes the set of electric buses, and $S_L$ denotes the set of distribution lines. A UTN is depicted by a graph $G_T = (S_M, S_A)$, where $S_M$ denotes the set of road nodes, representing origins, destinations, and intersections, and $S_A$ denotes the set of links, representing roadway segments. Furthermore, UTNs are expanded by introducing charging links and bypass links for CSs, as presented in [9]. Each vehicle travels from its origin $r \in S_r$ to its destination $s \in S_s$ through a path $k \in K_{rs}$ in UTNs. In the trip, each EV user will charge once at a CS on link $a \in S_C$, which connects to an electric bus $i \in S_N$ in PDNs. For the shared EVs controlled by AMoDS, their optimal routes and CS choices are determined by minimizing the total PTN operating cost:

$$\Gamma_{SO}: \quad \min \quad F_{PDN} + F_{UTN}^U$$
$$\text{s.t.} \quad PDN\ Constraints \quad (1)$$
$$UTN\ Constraints$$

where $F_{PDN}$ is the PDN operating cost, which is stated as:

$$F_{PDN} = \sum_{i \in S_N} \left[ \alpha_i \left( P_i^G \right)^2 + \beta_i P_i^G \right] + \lambda \sum_{k \in \pi(0)} P_{0k}^l + \gamma \sum_{j \in S_N} \Delta P_j^R \quad (2)$$

where the first term is the cost of traditional PDN generators, the second term is the electricity purchase cost from the utility grid, and the third term is the penalty cost for RDG power curtailments.

$F_{UTN}^U$ contains travel and charging costs of controllable EVs in UTNs, which are given by:

$$F_{UTN}^U = w \sum_{a \in S_A \cup S_C} x_a^U t_a + \sum_{a \in S_C} \lambda_j^a E_C x_a^U \quad (3)$$

where $\lambda_j^a$ is the locational marginal price (LMP) at the CS on charging link $a$, which is connected to the PDN bus $j \in S_N$.

The *PDN Constraints* include those of line flows and bus voltages, which are described by the DistFlow model with SOC relaxation [25], [26]:

$$P_{ij}^l + P_j^G + P_j^R - r_{ij}^l i_{ij}^l = \sum_{k \in \pi(j)} P_{jk}^l + P_j^D, \quad \forall l \in S_L \quad (4)$$

$$Q_{ij}^l + Q_j^G + Q_j^R - x_{ij}^l i_{ij}^l = \sum_{k \in \pi(j)} Q_{jk}^l + Q_j^D, \quad \forall l \in S_L \quad (5)$$

$$U_j = U_i - 2\left(r_{ij}^l P_{ij}^l + x_{ij}^l Q_{ij}^l\right) + \left(z_{ij}^l\right)^2 i_{ij}^l, \quad \forall l \in S_L \quad (6)$$

$$i_{ij}^l U_i \geq \left(P_{ij}^l\right)^2 + \left(Q_{ij}^l\right)^2, \quad \forall l \in S_L \quad (7)$$

$$i_{ij}^l \leq i_l^{\max}, \quad \forall l \in S_L \quad U_i^{\min} \leq U_i \leq U_i^{\max}, \quad \forall i \in S_N \quad (8)$$

$$P_i^{G,\min} \leq P_i^G \leq P_i^{G,\max}, \quad Q_i^{G,\min} \leq Q_i^G \leq Q_i^{G,\max}, \quad \forall i \in S_N \quad (9)$$

Constraints (4) and (5) show active and reactive power balances, respectively. Constraint (6) describes the voltage drop through lines. Constraint (7) is the SOC relaxation of apparent power constraints. Constraint (8) represents squared current and voltage magnitude limits. Constraint (9) represents active and reactive power generation limits.

The active power demands at bus $j$ include traditional load demands and EV charging demands, where:

$$P_j^D = P_j^T + P_j^E, \quad P_j^E = \left(x_a^U + x_a^L\right) E_C, \quad \forall j \in S_N \quad (10)$$

To balance the PDN operation, the active power of RDGs may be curtailed, which satisfies:

$$P_j^R = P_j^{R,\max} - \Delta P_j^R \quad (11)$$

The *UTN Constraints* characterize the traffic assignments. In the upper level, we use the SO principle-based static traffic assignment model to describe controllable EV flows. In AMoDS, there are two types of travel services, including regular and rebalance travels, which satisfy actual and prospective travel demands, respectively. Here, controllable EV flow constraints are stated as:

$$\sum_{k \in K_{rs}} f_k^{rs,A} = q_{rs}^A, \quad \forall r \in S_r, s \in S_s \quad (12)$$

$$\sum_{r \in S_r} \sum_{k \in K_{rs}} f_k^{rs,P} \geq q_s^P, \forall s \in S_s \quad \sum_{s \in S_s} \sum_{k \in K_{rs}} f_k^{rs,P} \leq q_r^P, \forall r \in S_r \quad (13)$$

$$x_a^{U,A} = \sum_{rs} \sum_{k \in K_{rs}} f_k^{rs,A} \delta_{ak}^{rs,A}, \quad x_a^{U,P} = \sum_{rs} \sum_{k \in K_{rs}} f_k^{rs,P} \delta_{ak}^{rs,P}, \quad \forall a \in S_A \cup S_C \quad (14)$$

$$x_a^U = x_a^{U,A} + x_a^{U,P}, \quad \forall a \in S_A \cup S_C \quad x_a^U + x_a^L \leq X_a, \quad \forall a \in S_C \quad (15)$$

$$t_a = t_a^0 \left[1 + 0.15 \left(\frac{x_a^U + x_a^L}{C_a}\right)^4\right], \quad \forall a \in S_A \cup S_C \quad (16)$$

Constraint (12) represents the traffic flow conservation. Constraint (13) represents the rebalance service limit. Constraint (14) represents the relationship between EV flow link and path for regular and rebalance travel services. Constraint (15) represents the total traffic flow on link $a$ and the capacity limit of charging links. Constraint (16) gives the travel time on link $a$ by the Bureau of Public Roads function. $\delta_{ak}^{rs,A} = 1$ ($\delta_{ak}^{rs,P} = 1$) if path $k$ satisfies the actual (prospective) travel demands from $r$ to $s$ passing link $a$; otherwise, $\delta_{ak}^{rs,A} = 0$ ($\delta_{ak}^{rs,P} = 0$).

The traffic flow $x_a^L$ is obtained by solving the lower-level problem and transmitted to the upper-level problem. Furthermore, uncontrollable GV driving behaviors as well as their impacts are considered in the lower-level model.

*2) Lower-level SUE Model:* The conventional UE model assumes that the travel cost can be evaluated accurately by each EV user. However, in practice, EV users only obtain partial information on congested links. Thus, travel costs are always biased, where EV users designate optimal paths with a certain level of uncertainty. Here, a logit model is used to describe stochastic choices for EV paths [27]. We assume travel cost estimation error on path $k \in K_{rs}$ follows identical and independent Gumbel distributions. Then, based on the logit model, the probability of selecting path $i$ for travel is:

$$p_i^{rs} = \frac{\exp\left(-\theta c_i^{rs}\right)}{\sum_{k \in K_{rs}} \exp\left(-\theta c_k^{rs}\right)}, \quad \forall r \in S_r, s \in S_s \quad (17)$$

where $\theta$ ($\theta \geq 0$) is the awareness coefficient of uncontrollable vehicle users regarding travel costs, and $c_k^{rs}$ is the travel cost of users on path $k \in K_{rs}$ under UE.

In [28], it turns out that (17) constitutes the Karush-Kuhn-Tucker condition of a stochastic traffic assignment model in UTNs under the UE principle. For PTNs, we further develop an SUE model considering uncontrollable EVs and GVs, which is given by:

$$\Gamma_{SUE}: \quad \min \quad F_{UTN}^L \\ \text{s.t.} \quad SUE \ Constraints \quad (18)$$

where $F_{UTN}^L$ is the operating cost of uncontrollable EVs and GVs in UTNs, which is stated as:

$$F_{UTN}^L = w \sum_{a \in S_A \cup S_C} \int_{x_a^U}^{x_a^U + x_a^L} t_a^0 \left[1 + 0.15 \left(\frac{x}{C_a}\right)^4\right] dx + \sum_{a \in S_C} \lambda_j^a E_C x_a^L \\ + \frac{1}{\theta} \sum_{rs} \sum_{k \in K_{rs}} \left(f_k^{rs,G} \ln f_k^{rs,G} + f_k^{rs,E} \ln f_k^{rs,E}\right) \quad (19)$$

The first term is the travel costs of uncontrollable EVs and GVs in UTNs. The second term is the charging costs of uncontrollable EVs. The third term quantifies the impact of path choice on travel costs. The *SUE Constraints* contain:

$$\sum_{k \in K_{rs}} f_k^{rs,G} = q_{rs}^G, \quad \sum_{k \in K_{rs}} f_k^{rs,E} = q_{rs}^E, \quad \forall r \in S_r, s \in S_s \quad (20)$$

$$x_a^{L,G} = \sum_{rs} \sum_{k \in K_{rs}} f_k^{rs,G} \delta_{ak}^{rs,G}, \quad x_a^{L,E} = \sum_{rs} \sum_{k \in K_{rs}} f_k^{rs,E} \delta_{ak}^{rs,E}, \quad \forall a \in S_A \cup S_C \quad (21)$$

$$x_a^L = x_a^{L,G} + x_a^{L,E}, \quad \forall a \in S_A \cup S_C \quad x_a^U + x_a^L \leq X_a, \quad \forall a \in S_C \quad (22)$$

Constraint (20) represents the traffic flow conservation for GVs and EVs. Constraint (21) represents the relationship between link and path traffic flows for EVs and GVs. Constraint (22) represents the total traffic flow on link $a$ and the capacity limit of charging links. $\delta_{ak}^{rs,G} = 1$ ($\delta_{ak}^{rs,E} = 1$) if path $k$ satisfies GV (EV) travel demands on link $a$; otherwise, $\delta_{ak}^{rs,G} = 0$ ($\delta_{ak}^{rs,E} = 0$).

In (19), $\theta$ is an input parameter of the lower-level SUE model. A larger $\theta$ indicates that the drivers can obtain more information on the PTN operational state (e.g., travel time on links and precise bus LMP at each CS). Thus, the travel cost perceived by drivers will be more accurate. Particularly, if $\theta \to +\infty$, the SUE model will be transformed into the conventional UE model, and the users will always select the path with the minimum travel cost. Therefore, $\theta$ reflects the influences of human cognitive behaviors on PTN operations.

## III. PROBABILISTIC POWER-TRAFFIC FLOW ANALYSIS

In this section, a probabilistic PTN flow method is developed to evaluate the impacts of various environmental and human uncertainties on PTN operations. The environmental uncertainties include the RDG power in PDNs and the UTN link capacities. The models for variable RDG power are provided in [29] and [30]. Normal distributions are used to depict the uncertain link capacity $C_a$ in (16) and (19). The human uncertainties include variable travel demands in (12) and (20), and the awareness coefficient $\theta$ for stochastic path choice behaviors in (19), which are also modeled by Normal distributions. These environmental and human uncertainties are regarded as input random variables for the PTF model. Hence, a probabilistic PTF analysis model is written in a compact form as follows:

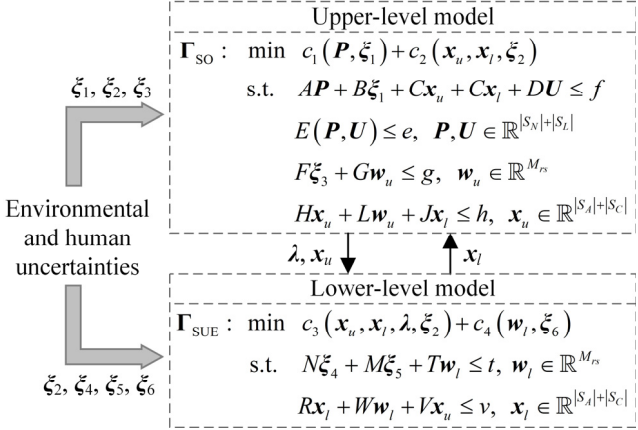

where $\xi_1$-$\xi_6$ are the input random variables, including the variable RDG power ($\xi_1$), degraded link capacities ($\xi_2$), variable travel demands in the upper-level model ($\xi_3$), variable travel demands in the lower-level model ($\xi_4$ for EVs and $\xi_5$ for GVs) and stochastic awareness coefficients ($\xi_6$). $\boldsymbol{P} := \{P_{ij}^I, P_i^G, \forall i, \forall l\}$; $\boldsymbol{U} := \{i_{ij}^l, U_j, \forall j, \forall l\}$; $\boldsymbol{x}_u := \{x_a^U, \forall a\}$; $\boldsymbol{x}_l := \{x_a^L, \forall a\}$; $\boldsymbol{w}_u := \{f_k^{rs,A}, f_k^{rs,P}, \forall r, \forall s\}$; $\boldsymbol{w}_l := \{f_k^{rs,G}, f_k^{rs,E}, \forall r, \forall s\}$; $\boldsymbol{\lambda}$ is the LMP vector; $E(\cdot)$ represents the SOC constraints; $M_{rs}$ is the number of O-D pairs; $A$, $B$, $C$, $F$, $J$, $H$, $L$, $N$, $M$, $T$, $R$, $W$, $V$ and $f$, $e$, $g$, $h$, $t$, $v$ are the coefficient matrices and vectors.

Given $N_{in}$ samples of input variables, the probabilistic PTF analysis is performed by the Monte Carlo simulations (MCS) with $N_{in}$ calculations of the bi-level PTF model. In each calculation, the upper-level model is converted into a mixed-integer second-order cone programming (MISOCP) problem [10], which is solved to offer the LMP vector $\boldsymbol{\lambda}$ and the controllable EV flow $\boldsymbol{x}_u$ to the lower-level problem. Then, the lower-level problem with a convex objective function and linear constraints is solved to obtain the traffic flow $\boldsymbol{x}_l$, which in turn is transmitted to the upper-level model. By solving the upper-level and lower-level models iteratively, the PTF results are derived when the calculation procedure is converged. Finally, the probabilistic results are obtained by performing $N_{in}$ calculations. The detailed PTF analysis process is presented in **Algorithm 1**, and the existence and uniqueness of solutions are elaborated on in [9].

| **Algorithm 1**: Probabilistic PTF analysis |
|---|
| 1: **Initialization:** Set the convergence tolerance $\varepsilon$ and the maximum iteration number $K_P$. Generate $N_{in}$ groups of input samples $\xi_1^{(i)}$-$\xi_6^{(i)}$ ($i=1,2,\ldots,N_{in}$). |
| 2: Establish the bi-level PTF models. |
| 3: **for** $i = 1$ to $N_{in}$ **do** |
| 4: Set initial traffic flows $\boldsymbol{x}_l$, $\boldsymbol{x}_u$ as zero, the initial value $\Delta^{(i)}$ ($\Delta^{(i)}>\varepsilon$), and the iteration number $j$ as 1. |
| 5: **while** $\Delta^{(i)}>\varepsilon$ or $j < K_P$ **do** |
| 6: Solve the upper-level model $\boldsymbol{\Gamma}_{SO}$ with the $i$th group of input sample $\xi_1^{(i)}, \xi_2^{(i)}, \xi_3^{(i)}$ using $\boldsymbol{x}_l$ to obtain the LMP $\boldsymbol{\lambda}^{(i)}$ and the traffic flow $\boldsymbol{x}_u^{(i)}$. |
| 7: Solve the lower-level model $\boldsymbol{\Gamma}_{SUE}$ with the $i$th group of input sample $\xi_2^{(i)}, \xi_4^{(i)}, \xi_5^{(i)}, \xi_6^{(i)}$ using $\boldsymbol{\lambda}^{(i)}$ and $\boldsymbol{x}_u^{(i)}$ to obtain the traffic flow $\boldsymbol{x}_l^{(i)}$. |
| 8: $\Delta^{(i)} = \|\boldsymbol{x}_l^{(i)} + \boldsymbol{x}_u^{(i)} - \boldsymbol{x}_l - \boldsymbol{x}_u\|$ |
| 9: **if** $\Delta^{(i)} \leqslant \varepsilon$ **then** |
| 10: Return the results of $\boldsymbol{\Gamma}_{SO}$ and $\boldsymbol{\Gamma}_{SUE}$ for $i$th MCS. |
| 11: **else if** $k = K_P$ **then** |
| 12: Report that the calculation is not converged. |
| 13: **else** |
| 14: $\boldsymbol{x}_l \leftarrow \boldsymbol{x}_l^{(i)}$, $\boldsymbol{x}_u \leftarrow \boldsymbol{x}_u^{(i)}$, $j=j+1$, and go to step 6. |
| 15: **end if** |
| 16: **end while** |
| 17: **end for** |
| 18: **Results:** Probabilistic PTF results. |

## IV. SENSITIVITY ANALYSES FOR UNCERTAINTIES

In this section, an RSA method is introduced to quantify the impacts of uncertainties on PTN operations. An LRA-based surrogate model is utilized to improve the computational efficiency of sensitivity analyses.

### A. Regional Sensitivity Index

Consider an evaluation model $Y=g(\boldsymbol{\xi})$ with $n$ inputs $\boldsymbol{\xi}=(\xi_1,\ldots,\xi_n)$, where $Y$ is the output variable with the probability density function (PDF) $f_Y(Y)$, and $\xi_i$ is $i$th input variable with PDF $f(\xi_i)$. To quantify the impacts of $\xi_i$ on the model output $Y$, the delta index $\delta_i$ is proposed in [31] as:

$$\delta_i = \frac{1}{2} E[s(\xi_i)] = \frac{1}{2} \int f(\xi_i) \int |f_Y(Y) - f_{Y|\xi_i}(Y)| \mathrm{d}Y \mathrm{d}\xi_i \quad (23)$$

where $E[\cdot]$ is the expectation function, $f_{Y|\xi_i}(Y)$ is the conditional PDF of $Y$ when $\xi_i$ is fixed as a constant, and $s(\xi_i)$ is the area difference between PDFs $f_Y(Y)$ and $f_{Y|\xi_i}(Y)$, as shown in Fig. 2.

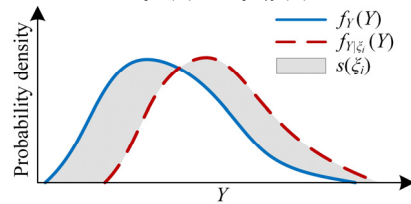

Fig. 2. Area difference between PDFs of $f_Y(Y)$ and $f_{Y|\xi_i}(Y)$.

Because $f_{Y,\xi_i}(Y, \xi_i) = f_{Y|\xi_i}(Y)f(\xi_i)$, (23) is rewritten as [32]:

$$\delta_i = \frac{1}{2}\iint |f_Y(Y)f(\xi_i) - f_{Y,\xi_i}(Y,\xi_i)| dYd\xi_i \quad (24)$$

where $f_{Y,\xi_i}(Y, \xi_i)$ is the joint PDF of $Y$ and $\xi_i$.

Further, a Copula function [33] is introduced to represent (24); thus, the cumulative distribution function (CDF) of $Y$ and $\xi_i$ is given by:

$$F_{Y,\xi_i}(Y,\xi_i) = C(F_Y(Y), F(\xi_i)) \quad (25)$$

where $F_Y(Y)$ and $F(\xi_i)$ are the CDFs of $Y$ and $\xi_i$, respectively.

Accordingly, the Copula density function $c(F_Y(Y), F(\xi_i))$ is:

$$\begin{aligned}c(F_Y(Y), F(\xi_i)) &= \frac{\partial^2 C(F_Y(Y), F(\xi_i))}{\partial F_Y(Y)\partial F(\xi_i)}\\ &= \frac{\partial^2 F_{Y,\xi_i}(Y,\xi_i)}{\partial Y \partial \xi_i}\frac{\partial Y}{\partial F_Y(Y)}\frac{\partial \xi_i}{\partial F(\xi_i)}\\ &= \frac{f_{Y,\xi_i}(Y,\xi_i)}{f_Y(Y)f(\xi_i)}\end{aligned} \quad (26)$$

Hence, (24) is expressed as:

$$\begin{aligned}\delta_i &= \frac{1}{2}\int_0^1\int_0^1 |f_Y(Y)f(\xi_i) - c(F_Y(Y),F(\xi_i))f_Y(Y)f(\xi_i)| dYd\xi_i\\ &= \frac{1}{2}\int_0^1\int_0^1 |1 - c(F_Y(Y),F(\xi_i))| dF_Y(Y)dF(\xi_i)\end{aligned} \quad (27)$$

To further evaluate the impacts of distribution ranges $[F^{-1}(q_1), F^{-1}(q_2)]$ ($q_1$ and $q_2$ are quantiles), a regional sensitivity index $\gamma_i$ is defined as:

$$\gamma_i = \frac{1}{2\delta_i(q_2-q_1)}\int_{q_1}^{q_2}\int_0^1 |1 - c(F_Y(Y),F(\xi_i))| dF_Y(Y)dF(\xi_i) \quad (28)$$

Fig. 3 gives an intuitive description of the $\delta$ and $\gamma_i$ indices. The $\delta$ index in (27) quantifies the impact of $\xi_i$ on $Y$ considering the variation range of $\xi_i$. The $\delta$ index, which is called the global sensitivity index, is used to identify critical uncertainties by a sensitivity ranking. The regional sensitivity index $\gamma_i$ in (28) quantifies the impacts of certain ranges of $\xi_i$ on $Y$, which provides a microcosmic perspective for the uncertainty analysis.

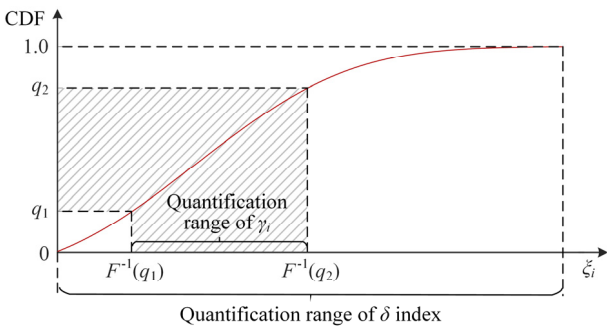

Fig. 3. Quantification ranges of $\delta$ and $\gamma_i$.

### B. Efficient Calculation Method of Sensitivity Indices

Given the samples of inputs $\boldsymbol{\xi}^S$, the samples of output $\boldsymbol{Y}^S$ can be obtained by MCS. Then, the sensitivity indices are calculated by estimating the Copula functions and CDFs in (27) and (28) using kernel density estimation [34]. In this paper, the output samples are obtained by the probabilistic PTF analysis, which is time-consuming for solving a large number of bi-level PTF models. To improve the computational efficiency, a surrogate model is established using a low-rank approximation (LRA) to replace the original PTF model, which is stated as:

$$M(\boldsymbol{\xi}) = \sum_{w=1}^{r_{max}} b_w z_w(\boldsymbol{\xi}) \quad (29)$$

where $r_{max}$ is the maximum rank, and $z_w(\boldsymbol{\xi})$ comprises a group of rank-one components, stated as:

$$z_w(\boldsymbol{\xi}) = \prod_{i=1}^{n}\left(\sum_{d=0}^{D_{i,max}} a_{d,w}^{(i)}\varphi_d^{(i)}(\xi_i)\right) \quad (30)$$

where $d$ is the degree of univariate polynomials $\varphi_d^{(i)}(\xi_i)$, and $D_{i,max}$ is the maximum of $d$ for each input variable $\xi_i$.

The LRA-based surrogated model is a tensor representation of the original model using orthonormal polynomials [35]. The procedure for calculating sensitivity indices via LRA is given in **Algorithm 2**.

| **Algorithm 2**: Calculate sensitivity indices by LRA model |
|---|
| 1: **Initialization:** Collect the PTF analysis results, including the samples of inputs $\boldsymbol{\xi}^S$ and output $\boldsymbol{Y}^S$. |
| 2: Establish the LRA-based surrogated model $M(\boldsymbol{\xi})$. |
| 3: Generate $N_{LRA}$ samples of $Y$ using $M(\boldsymbol{\xi})$. |
| 4: **for** $i = 1$ to $n$ **do** |
| 5:     Obtain CDFs of $\xi_i$ and $Y$. |
| 6:     Derive the Copula density $c(F_Y(Y), F(\xi_i))$. |
| 7:     Calculate the $\delta_i$ index by numerical integration. |
| 8:     Set quantiles $q_1$ and $q_2$, and calculate the $\gamma_i$ index by numerical integration. |
| 9: **end for** |
| 10: **Results:** Sensitivity indices of $\xi_1,\ldots,\xi_n$ |

## V. CASE STUDIES

### A. PTN System Description

The effectiveness of the proposed PTF analysis method is verified by a PTN consisting of a 21-bus PDN and a 20-node UTN, as shown in Fig. 4. For the PDN, there are six RDGs, including three wind turbines (i.e., $WT_1$, $WT_2$, $WT_3$) and three photovoltaic cells (i.e., $PV_1$, $PV_2$, $PV_3$). The PDN base power is 40 MVA. The base active and reactive power demands at each bus $j$, i.e., $P_j^T$ and $Q_j^T$, are 0.02 p.u. and 0.01 p.u., respectively. Four traditional DGs are integrated into the PDN at buses 7, 10, 11, and 14, respectively. The line and generator parameters are from [36]. The electricity price $\rho$ at the utility grid (UG) is 150 \$/MWh. The voltage magnitude at the slack bus $U_0$ is 1.04 p.u. and $U_i^{min}$ and $U_i^{max}$ are 0.9025 p.u. and 1.1025 p.u., respectively.

For the UTN, eight CSs are located at nodes 13-20, which are connected to PDN buses 2, 9, 7, 10, 11, 12, 18, and 20, respectively. The electricity price is set as 100 \$/MWh at each CS for the EVs controlled by AMoDS. For the charging link $a \in S_C$, the free travel time $t_a^0$ and the capacity $C_a$ are 20 min and 15 p.u., respectively. For the bypass link of CSs, the free travel time $t_a^0$ is zero. Also, the UTN contains four kinds of links, as shown in Table I. The O-D pairs and the actual traffic demands of controllable EVs (denoted as $EV_c$), GVs, and uncontrollable

EVs (denoted as EV$_{uc}$) are given in Table II. The predictive traffic demand $q_s^P$ at rebalance destination nodes 6, 10, and 12 is 4 p.u., and the available EV limit $q_r^P$ at rebalance origin nodes 1, 3, and 4 is 20 p.u. The predictive traffic demands are satisfied by the rebalance traffic flow through the AMoDS control. The base value of traffic flow is 10 vehicles/hour. The travel time monetary cost coefficient $\omega$ is 10 \$/h, and the charging demand $E_C$ of each EV is 20 kWh.

There are 34 input variables, including $\xi_1$-$\xi_3$ for the power outputs of WT$_1$-WT$_3$, $\xi_4$-$\xi_6$ for the power outputs of PV$_1$-PV$_3$, $\xi_7$-$\xi_{17}$ for the travel demands of GVs, $\xi_{18}$-$\xi_{28}$ for the travel demands of EVs, $\xi_{29}$-$\xi_{33}$ for the capacities of roads and charging links, and $\xi_{34}$ for the awareness coefficient of uncontrollable vehicle users. The power output PDFs of six RDGs are found in Fig. 5. Normal distribution functions with a coefficient of variation of 10% are used to depict other environmental and human uncertainties. For the PTF analysis, the upper-level MISOCP model is solved by CPLEX [37], and the lower-level nonlinear model is solved by Baron [38]. The UQlab is used to build the LRA-based surrogate model for RSA [39]. All simulations are implemented with Matlab 2018a on a PC with Intel Core i5-8500 3.00GHz CPU and 8GB memory.

TABLE I
PARAMETERS OF LINKS

| Parameters | Type-1 | Type-2 | Type-3 | Type-4 |
|---|---|---|---|---|
| $C_a$(p.u.) | 25 | 25 | 20 | 15 |
| $t_a^0$(min) | 5 | 8 | 5 | 7 |

TABLE II
O-D PAIRS AND ACTUAL TRAVEL DEMANDS (P.U.)

| O-D pair | EV$_c$ $q_{rs}^A$ | GV $q_{rs}^G$ | EV$_{uc}$ $q_{rs}^E$ | O-D pair | EV$_c$ $q_{rs}^A$ | GV $q_{rs}^G$ | EV$_{uc}$ $q_{rs}^E$ |
|---|---|---|---|---|---|---|---|
| 1-6 | 1.5 | 7.5 | 1.5 | 3-6 | 1.5 | 7.5 | 1.5 |
| 1-10 | 1 | 15 | 1 | 3-10 | 1 | 12.5 | 1 |
| 1-11 | 1.5 | 10 | 1.5 | 3-11 | 1.5 | 10 | 1.5 |
| 1-12 | 1 | 10 | 1 | 3-12 | 1 | 12.5 | 1 |
| 4-9 | 1.5 | 12.5 | 1.5 | 4-10 | 1.5 | 10 | 1.5 |
| 4-12 | 1.5 | 10 | 1.5 | | | | |

### B. Probabilistic Power-Traffic Flow Analysis

In this section, we obtain the probabilistic PTF results by MCS. The sample size is 2000, and the Sobol sequence is used to generate input samples. Then, the probability distributions and statistical indices of outputs are obtained to analyze the impacts of uncertainties.

Figs. 6 and 7 show the voltage magnitude probability distribution at each PDN bus and the UTN traffic flow, respectively. The total traffic flow contains controllable EV, uncontrollable EV, and GV flows. For the PDN, the variations of voltage magnitudes at buses 1, 3, 6, and 8 are significantly large, while the voltage magnitudes at buses 2, 4, 5, 7, 14, 17, 18, and 20 fluctuate slightly around their means. For the UTN, the traffic flow variations on links 1→13, 3→15, 4→14 and 4→16 are larger than those on other links. The maximum traffic flow varies from 7.8 p.u. to 85.9 p.u. on link 4→14. The tail distribution of traffic flows in Fig. 7 implies that traffic flows may vary dramatically in extreme scenarios, which should alert UTN operators to avoid serious congestion or extremely unbalanced traffic flows.

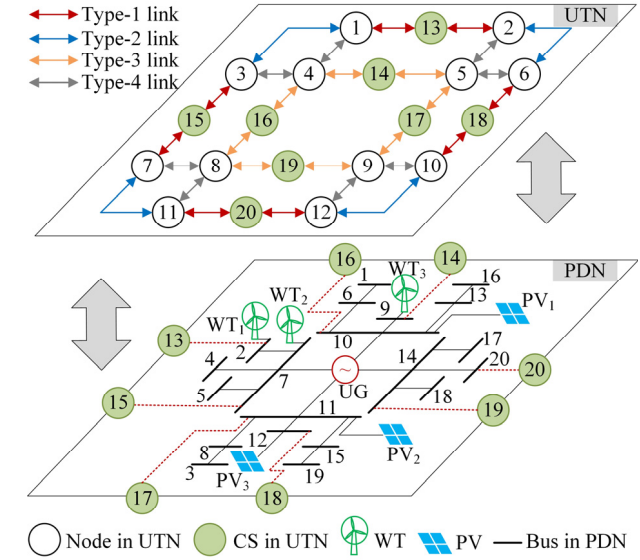

Fig. 4. PTN consisting of PDN and UTN.

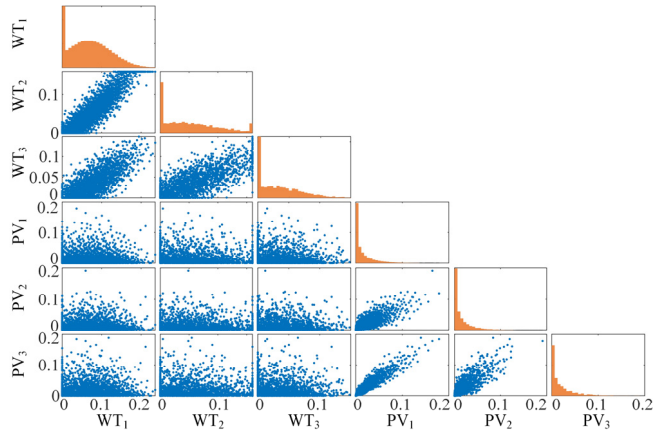

Fig. 5. Distribution of RDG power output (p.u.).

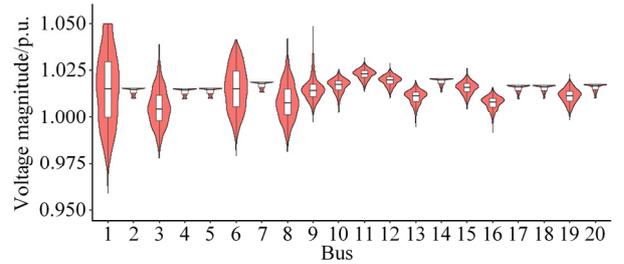

Fig. 6. Voltage magnitude distribution at each PDN bus.

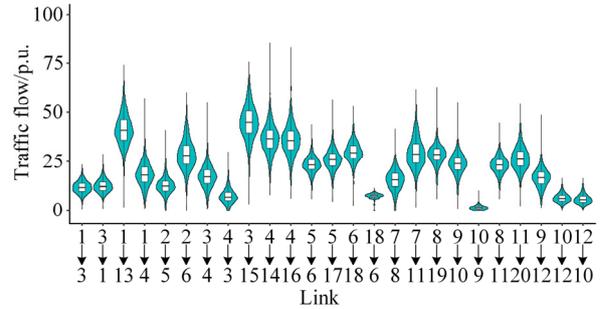

Fig. 7. Traffic flow distribution on UTN link.

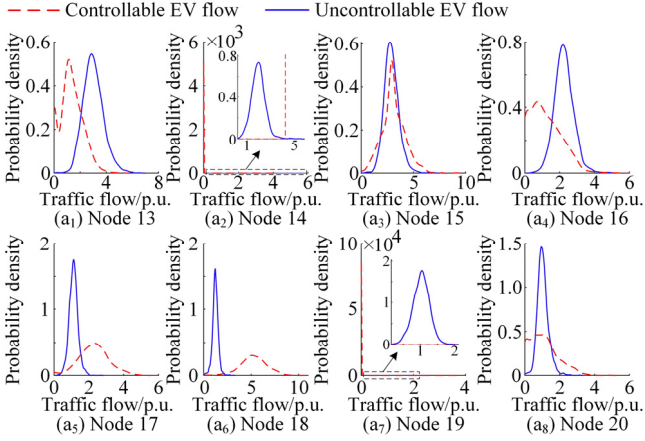

Fig. 8. PDFs of controllable and uncontrollable EV flows through each CS.

TABLE III
MEAN AND VARIANCE OF CONTROLLABLE AND UNCONTROLLABLE EV FLOWS THROUGH EACH CS

| CS node | Controllable EV flow | | Uncontrollable EV flow | |
|---|---|---|---|---|
| | Mean/p.u. | Variance/p.u. | Mean/p.u. | Variance/p.u. |
| 13 | 1.32 | 0.74 | 3.00 | 0.59 |
| 14 | 0.31 | 0.45 | 1.98 | 0.35 |
| 15 | 3.08 | 1.25 | 2.90 | 0.51 |
| 16 | 1.17 | 0.81 | 2.24 | 0.30 |
| 17 | 2.27 | 0.75 | 1.12 | 0.06 |
| 18 | 5.23 | 1.96 | 1.23 | 0.09 |
| 19 | 0.07 | 0.09 | 1.03 | 0.06 |
| 20 | 1.06 | 0.67 | 1.04 | 0.10 |
| Total | 14.51 | 6.71 | 14.51 | 2.07 |

Fig. 8 gives the PDFs of controllable and uncontrollable EV flows through each CS, and Table III gives the means and variances of EV flows. The controllable EV flow has a wider variation range than the uncontrollable one at the same CS. The total variance of the controllable EV flow is three times larger than that of the uncontrollable EV flow, indicating that the uncertainties have a remarkable impact on the CS choices of controllable EVs. The means of controllable and uncontrollable EV flows through each CS are different, but their total means are the same. It is because the travel demand of controllable EVs is the same as that of uncontrollable EVs, as shown in Table II.

Based on PTF results, the correlation between electricity prices and EV flows is analyzed. The correlation coefficient between the LMP $\lambda_i$ at CS node $i$ ($i$ = 13, 14, …, 20) and the controllable EV flow $x_{c,i}$ is illustrated in Fig. 9(a). The correlation coefficient between $\lambda_i$ and the uncontrollable EV flow $x_{uc,i}$ is illustrated in Fig. 9(b). The LMPs at different CSs are strongly correlated when correlation coefficients are larger than 0.85. The mean values of LMPs at different CSs are similar, as given in Table IV, indicating that the interaction between PDNs and UTNs could result in similar electricity prices at different CSs. In contrast, LMP variances are different, indicating that the uncertainties could have various impacts on CS operations.

The correlation among uncontrollable EV flows is stronger than that among controllable EV flows. The maximum correlation coefficient among uncontrollable EV flows is 0.88, while that among controllable EV flows is 0.36. This means that the charging choice behavior is more correlated under the UE principle. For example, two CSs at nodes 14 and 17 are on the path 1→4→14→5→17→9→10. The EV users would choose the CS at node 14 or 17 for charging when traveling through this path. To obey the UE principle, the charging choices for the two CSs will be closely related, making uncontrollable EV flows through the two CSs strongly correlated. Comparatively, controllable EVs choose CSs to minimize the total cost under the control of AMoDS. The CS scheduling may be different in various scenarios, so that the controllable EV flows through different CSs are weakly correlated.

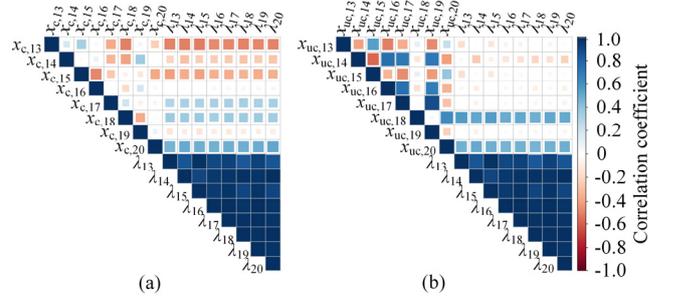

Fig. 9. Correlation coefficients among LMP $\lambda_i$ at the $i$th CS node and (a) controllable EV flow $x_{c,i}$ through CS node $i$, (b) uncontrollable EV flow $x_{uc,i}$ through CS node $i$.

TABLE IV
MEAN AND VARIANCE OF LMP AT EACH CS NODE

| CS node in UTN | Connecting bus in PDN | Mean ($/MWh) | Variance ($^2$/MWh$^2$) |
|---|---|---|---|
| 13 | 2 | 147.89 | 171.32 |
| 14 | 9 | 149.48 | 62.50 |
| 15 | 7 | 147.43 | 99.07 |
| 16 | 10 | 147.01 | 54.52 |
| 17 | 11 | 146.48 | 49.36 |
| 18 | 12 | 145.04 | 38.67 |
| 19 | 14 | 147.78 | 50.21 |
| 20 | 20 | 147.67 | 43.47 |

*C. RSA for Uncertainties*

*1) Identifying critical uncertainties:*

We select six output variables, including the voltage magnitudes at bus 10 ($V_{10}$) and bus 20 ($V_{20}$), the traffic flows on link 1→4 ($x_{1\to4}$) and link 7→11 ($x_{7\to11}$), and the traffic flows through CSs at node 16 ($x_{16}$) and node 18 ($x_{18}$). For each output variable, the $\delta$ indices of all input variables $\xi_i$ ($i$ = 1,2,…,34) are calculated to determine the importance ranking of inputs, as shown in Fig. 10. For the PDN, the variable output power of PV$_1$ ($\xi_4$) and WT$_1$ ($\xi_1$) are the most important inputs for $V_{20}$ and $V_{10}$, respectively. The capacity uncertainty of Type-3 link ($\xi_{31}$) also has significant impacts, and it is ranked 3rd and 2nd for $V_{10}$ and $V_{20}$, respectively. It indicates that the link capacity in the UTN affects the operational state of the PDN.

For the UTN, WT$_1$ ($\xi_1$) is ranked 1st for $x_{18}$ and PV$_1$ ($\xi_4$) is ranked 1st for $x_{16}$. Comparatively, the RDGs are ranked low for $x_{1\to4}$ and $x_{7\to11}$. This indicates that the variable RDG power mainly affects the EV flow through CSs and does not affect the traffic flows on other links in the UTN. On the contrary, variable link capacities (i.e., $\xi_{29}$-$\xi_{32}$) have significant impacts on traffic flows without CSs (i.e., $x_{1\to4}$ and $x_{7\to11}$) as well as the traffic flows through CS nodes (i.e., $x_{16}$ and $x_{18}$). Also, the $\delta$ index of awareness coefficient $\theta$ quantifies the impact of user's path choice behaviors, whose importance is moderate overall.

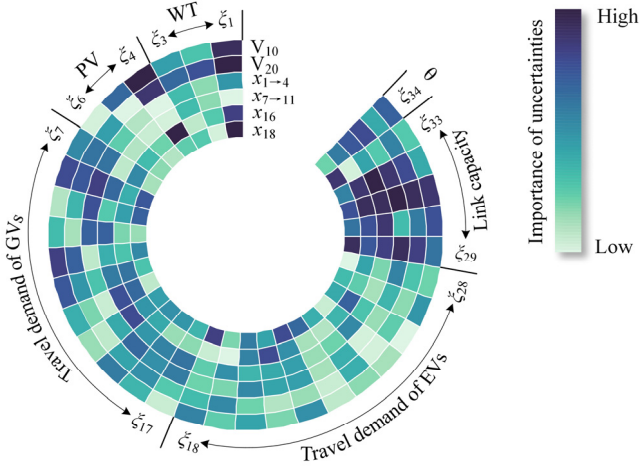

Fig. 10. Ranking of important uncertainties for output variables.

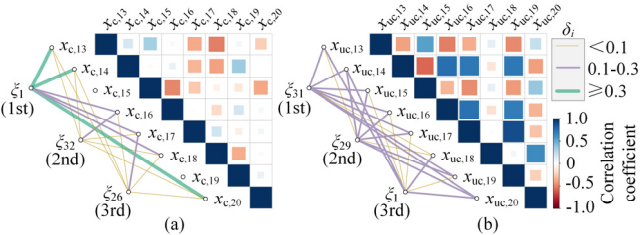

Fig. 11. Three most critical uncertainties for (a) controllable EV flow $x_{c,i}$ and (b) uncontrollable EV flow $x_{uc,i}$.

Further, we select the controllable EV flow $x_{c,i}$ ($i = 13, 14,…,20$) and the uncontrollable EV flow $x_{uc,i}$ ($i = 13,14,…,20$) as output variables. Based on the mean of $\delta$ indices, the three most critical uncertainties for $x_{c,i}$ and $x_{uc,i}$ are identified, as shown in Figs. 11(a) and 11(b), respectively. The identified critical uncertainties include the variable power of WT$_1$ ($\xi_1$), link capacities (i.e., $\xi_{29}$, $\xi_{31}$, $\xi_{32}$), and EV travel demands ($\xi_{26}$). Meanwhile, the mean $\delta$ index of $\xi_1$ for controllable EV flows is larger than that for uncontrollable EV flows. This indicates that the impact of the variable power of WT$_1$ ($\xi_1$) on controllable EV flows is greater than its impact on uncontrollable EV flows. The EV travel demand in the O-D pair 3-11 ($\xi_{26}$) is identified as one of the critical uncertainties for controllable EV flows. In comparison, the impact of the EV travel demand on uncontrollable EV flows is minor in this case.

*2) Impacts of distribution range of uncertainties:*

We select the CS traffic flow at node 18 ($x_{18}$) and the voltage magnitude at bus 10 ($V_{10}$) as the outputs. We select the WT$_1$ power ($\xi_1$) as the critical input variable and use RSA to analyze the impacts of distribution ranges on the outputs. Figs. 12(a) and 13(a) show the respective results of regional sensitivity indices of $\xi_1$ for $x_{18}$ and $V_{10}$ for arbitrary quantile intervals [$q_1$, $q_2$]. The CDF of $\xi_1$ and its distribution ranges are presented in Figs. 12(b) and 13(b).

In Fig. 12, the regional sensitivity indices $\gamma_1(x_{18})$ at points A and B are 2.02 and 0.57, respectively. The quantile interval at point A is [0.8, 1.0], which corresponds to the distribution range II of $\xi_1$, and the quantile interval at point B is [0.4, 0.6], which corresponds to the distribution range I of $\xi_1$. Although the quantile intervals at points A and B have the same length, the regional sensitivity index at point A is larger than that at point B, which indicates that $\xi_1$ varying in range II makes a larger influence on $x_{18}$, as compared with that in range I. In Fig. 13, both points C and D have relatively large regional sensitivity indices, which indicates that $\xi_1$ varying in ranges II and III will have significant impacts on $V_{10}$.

The RSA results present more information to the operator for monitoring the crucial ranges of uncertainties. For example, the closer points to point A have large regional sensitivity indices. Hence, we should pay more attention to $\xi_1$ variations in range II, because a large fluctuation of $x_{18}$ would cause severe traffic congestion at the CS of node 18.

*3) Comparison of RSA with traditional GSA methods:*

We compare the proposed RSA method with the traditional global sensitivity analysis (GSA) methods，including the Sobol' and Borgonovo $\delta$ methods. By the LRA-based surrogated model, the computation time for regional sensitivity indices is reduced by 90% as compared with that using MCS. The critical uncertainties identified by the three methods are given in Table V, when $V_{10}$ is selected as the output. The RSA method is computationally more efficient than the Borgonovo $\delta$ method, but it needs a longer computation time than the Sobol' method. The reason is that the proposed method requires fewer samples than that of the Borgonovo $\delta$ method, but the proposed sensitivity calculation method still relies on numerical integration while the Sobol' index is calculated via analytical solutions. The three methods obtain the same identification results, which verifies the effectiveness of the proposed RSA method. It is worth noting that the RSA method can quantify the impacts of distribution ranges, which cannot be achieved by other GSA methods.

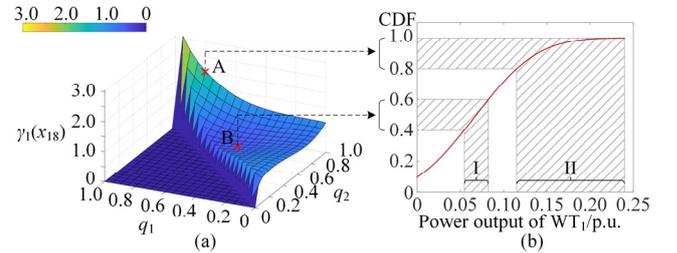

Fig. 12. (a) Regional sensitivity index $\gamma_1(x_{18})$ of $\xi_1$ for output variable $x_{18}$ in quantile interval [$q_1$, $q_2$]; (b) CDF and distribution ranges of quantile intervals [0.4, 0.6] and [0.8, 1.0].

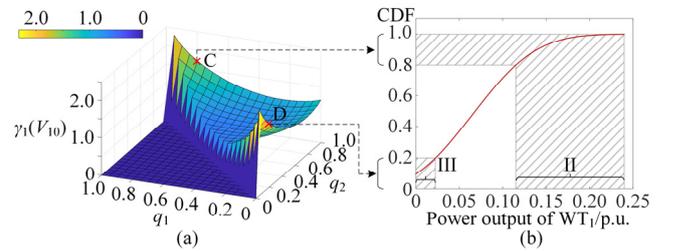

Fig. 13. (a) Regional sensitivity index $\gamma_1(V_{10})$ of $\xi_1$ for output variable $V_{10}$ in quantile interval [$q_1$, $q_2$]; (b) CDF and distribution ranges of quantile intervals [0, 0.2] and [0.8, 1.0].

TABLE V
COMPUTATION TIME AND IDENTIFICATION RESULTS OF THREE METHODS

| Method | Time(s) | Six most critical uncertainties |
|---|---|---|
| Borgonovo $\delta$ | 326.38 | $\xi_4$, $\xi_1$, $\xi_{31}$, $\xi_{11}$, $\xi_8$, $\xi_{30}$ |
| Sobol' | 9.60 | $\xi_4$, $\xi_1$, $\xi_{31}$, $\xi_{30}$, $\xi_8$, $\xi_{11}$ |
| RSA | 88.24 | $\xi_4$, $\xi_1$, $\xi_{31}$, $\xi_{11}$, $\xi_8$, $\xi_{30}$ |

## VI. CONCLUSION

This paper proposes a probabilistic method for analyzing PTNs with AMoDS under environmental and human uncertainties. The simulation results show that:
1. PTNs feature stochastic characteristics for engaging various uncertainties. The PTN operator should pay additional attention to those state variables which are varying in a large range or having a distinct tail distribution, which may cause system states to exceed operational limits.
2. In a CS, the variance of controllable EV flows is larger than that of uncontrollable EV flows. The uncertainties make a larger impact on CS choices for controllable EVs in PTN operations.
3. The proposed RSA method effectively identifies the critical uncertain inputs and quantifies the impacts of their distribution ranges on the concerned output. It provides a new perspective for analyzing the impacts of uncertainties on PTN operations.